\documentclass[12pt]{article}

\newtheorem{lemma}{Lemma}[section]

\newcommand {\dn}[1] {\boldsymbol #1}

\newcommand{\qed}{\nobreak \ifvmode \relax \else
      \ifdim\lastskip<1.5em \hskip-\lastskip
      \hskip1.5em plus0em minus0.5em \fi \nobreak
      \vrule height0.75em width0.5em depth0.25em\fi}

\usepackage[sort&compress]{natbib}
\usepackage{latexsym,rotating,amsfonts,amsmath,amssymb,float}
\usepackage{multirow}
\usepackage{comment}
\usepackage{epsf,epsfig}
\usepackage{subfigure}
\usepackage{psfrag,fancybox}
\usepackage[dvips]{color}
\usepackage{graphicx}
\usepackage{epstopdf}
\usepackage{float}
\usepackage{nomencl}
\usepackage{epsf,epsfig}
\usepackage{fullpage,subfigure}
\usepackage{psfrag,fancybox}
\usepackage[dvips]{color}
\usepackage{graphicx}
\usepackage{natbib}
\usepackage{amsmath}

\newcommand{\pref}[1]{%
    \ref{#1} \ifnum\count0=\pageref{#1}\relax%
    \else (page \pageref{#1})\fi}

\newcommand{\eref}[1]{%
        \ref{#1}\ifnum\count0=\pageref{#1}\relax%
        \else {, p.\pageref{#1}}\fi}

\newlength{\labwidth}
\newcommand{\bs}{\boldsymbol }
\newcommand{\noi}{\noindent }
\newcommand{\mo}{ m_0}
\newcommand{\mi}{ m_1}
\newcommand{\mii}{m_i}

\DeclareMathAlphabet\mathbfcal{OMS}{cmsy}{b}{n}

\newcommand{\cites}[1]{\citeauthor*{#1}'s \citeyearpar{#1}} 

\begin{document}

\title{\textbf{\sf Thermodynamic assessment of probability distribution divergencies and Bayesian model comparison }}

\author{Silia Vitoratou and Ioannis Ntzoufras \\
				\small \it Department of Statistics,
				\small \it Athens University of Economics and Business, Greece}

\date{15 October 2013}

\maketitle

\abstract{
Within path sampling framework, we show that probability distribution divergences, such as the Chernoff information, can be estimated via thermodynamic integration. The Boltzmann-Gibbs distribution pertaining to different Hamiltonians is implemented to derive tempered transitions along the path, linking the distributions of interest at the endpoints. Under this perspective, a geometric approach is feasible, which prompts intuition and facilitates tuning the error sources. Additionally, there are direct applications in Bayesian model evaluation. Existing marginal likelihood and Bayes factor estimators are reviewed here along with their stepping-stone sampling analogues. New estimators are presented and the use of compound paths is introduced.
}

 \bigskip


\section{Introduction}
\label{Intro}

The idea of using tempered transitions has gained increased attention in Bayesian statistics as
a method to improve the efficiency of Markov chain Monte Carlo (MCMC) algorithms in terms of exploring the target posterior distribution.
Sophisticated methods such as the Metropolis-coupled MCMC (\citealp{Geyer91}), the simulated
tempering (\citealp{MarPar92}; \citealp{GeTh95}) and the annealed sampling \citep{Neal96,Neal01} incorporate transitions
to overcome the slow mixing of the MCMC algorithms in multi-modal densities;
see \cite{Behetal12} for an insightful review.

Here, we focus on the ideas of path sampling \citep{GelMeng94,GelMeng98} where tempered transitions are employed
in order to estimate the ratio of two intractable normalizing constants. In particular,
let  $q_0(\bs{\theta})$ and $q_1(\bs{\theta})$ be two unnormalized densities
and  $z_0$, $z_1$ be their normalizing constants  leading to
\begin{equation}\label{zt}
p_t(\bs{\theta})=\frac{q_t(\bs{\theta})}{z_t },\mbox{~~where~} z_t=\int_{\bs{\theta}} q_t(\bs{\theta})\,d\bs{\theta},
~~\mbox{for~}t=0,1.
\end{equation}
\cites{GelMeng98} method is based on the construction of a continuous and differentiable \emph{path}
$q_t(\bs{\theta})=h(q_1,q_0,t)$ which is used to estimate the ratio of normalizing constants
$\lambda=z_1/z_0$ via the \emph{thermodynamic integration} (TI)  identity
\begin{eqnarray}\label{Identity_TI}
\log \lambda =
 \int_0^1 \int_{\bs{\theta}} \frac{d \,\log \,q_t(\bs{\theta})}{dt}\,p_t(\bs{\theta})\, d \bs{\theta}\,dt=\int_0^1 \!
 E_{p_t}\! \big\{U(\bs{\theta})\big\} dt,
\end{eqnarray}
where
$U(\bs{\theta}) = \frac{d \,\log \,q_t(\bs{\theta})}{dt}$
and
$E_{p_t}\! \big\{U(\bs{\theta})\big\}$ stands for the expectation over the sampling distribution $p_t(\bs{\theta})$.
The scalar $t \in [0,1]$ is often referred to as the  \emph{temperature} parameter, since  the TI
has its origins in thermodynamics and specifically in the calculation of the difference in \emph{free energy} of a system;
for details see in \citet[Section 6.2]{Neal93}. It occurs that the ideas of the thermodynamics
have important applications on a variety of scientific fields,
such as statistics, physics, chemistry, biology and computer science (machine learning, pattern recognition) among others.
As \cite{GelMeng98} denote, methods related to the TI have been developed by researchers
from different disciplines working independently and in parallel;
see, for instance, in \cite{Frenkel86}, \cite{Binder86} and \cite{Ogata89}.

A straightforward application of the path sampling refers to Bayesian model comparison.
In particular, expressions for the Bayes factor (BF, \citealp{kas:raf95}) and the marginal likelihood
that employ tempered transitions have been developed by \cite{LarPhil06}, \cite{FrielPet08}, \cite{Xieetal11} and \cite{Fanetal11}.
Additionally,  \cite{FrielPet08}, \cite{CalGir09}, \cite{Lefetal2010} and \cite{Behetal12}, under different motivations and scopes,
outline the close relationship between the thermodynamic integration and the relative entropy, best known in statistics as
the Kullback-Leibler divergence (KL; \citealp{KuLeib51}).

All these studies, are based on specific geometric paths \citep{Neal93} of the  form
\begin{equation}\label{Path_geom}
 q_t(\bs{\theta})=q_1(\bs{\theta})^t q_0(\bs{\theta})^{1-t},
\end{equation}
 for specific choices of  $q_0(\bs{\theta})$ and   $q_1(\bs{\theta})$.
 For example, \cite{FrielPet08} have used $q_t(\bs{\theta})= f( \dn{y} | \dn{\theta} )^{t} f( \dn{\theta} )$
 and therefore setting the unnormalized posterior as $q_1$ and the prior as $q_0$.
Here, we focus on the general case of geometric paths (\ref{Path_geom}) for any choice of $q_1$ and $q_0$.
For any geometric path, (\ref{Identity_TI}) is written as
\begin{eqnarray}\label{Identity_TIgeom}
\log\lambda= \int_0^1 \int_{\bs{\theta}} \log \frac{q_1(\bs{\theta})}{q_0(\bs{\theta})}\, p_t(\bs{\theta})\, d\bs{\theta}\, dt.
\end{eqnarray}
since
$U(\dn{\theta}) = \log q_1(\bs{\theta}) - \log q_0(\bs{\theta})$~.

We focus on (\ref{Identity_TIgeom}) in order to study the connection between path sampling and entropy measures.
In particular, we examine what happens for specific values of $t\in(0,1)$ and we describe the  mechanism which eventually produces
the relative entropy at the initial ($t=1$) and at the final ($t=0$) state,
as originally discussed by \cite{FrielPet08} and \cite{Lefetal2010}.
We demonstrate that (\ref{Identity_TIgeom}) can be used to compute the Chernoff information \citep{Chernoff52} as a byproduct
of the path sampling procedure, which is, otherwise, a rigorous and troublesome procedure especially in multidimensional problems.
Other  entropy measures can be subsequently derived, such as the \emph{Bhattacharyya distance} \citep{Bhatt43} and R\'{e}nyi's relative entropy  \citep{Renyi61}.

Based on our findings with regard to the uncertainty at the intermediated points,
we further examine and geometrically represent the structure of the thermodynamic integration.
This assists us to understand the path sampling estimators in terms of error.
In particular, can identify when high path-related uncertainty or large discretisation error
appears and reduce it by either adopting a more efficient (in terms of error) path
or tempering schedule.

Finally, we restrict attention on the most popular implementation of TI estimation: Bayesian model evaluation.
We further consider an alternative approach based on the stepping-stone identity introduced by \cite{Xieetal11} and \cite{Fanetal11}.
Then, we overview existing marginal likelihood estimators based on the two alternative approaches
(thermodynamic and stepping-stone)
by presenting  recently developed TI based marginal likelihood estimators \citep{LarPhil06,FrielPet08,Lefetal2010}
and their corresponding stepping-stone ones \citep{Xieetal11,Fanetal11} based on same paths.
Any blanks in the list of previously reported estimators based on the two different approaches
are filled in by introducing new estimators using a identity-path selection rationality.
We further discuss the implementation of the two alternative approaches
in the direct Bayes factor estimation and we introduce compound paths
which can be used to efficiently switch between
competing models of different dimension located  at the endpoints of the path.
The paper closes with an illustration of our methods and estimators
in a common regression example
(previously used by \citealp{FrielPet08} and \citealp{Lefetal2010} for marginal likelihood estimation)
and in a latent-trait model implementation using a simulated dataset.

\section{Entropy measures and path sampling}

In Statistics, entropy is used as a measure of uncertainty which, unlike the variance, does not
depend on the actual values of a random variable $\bs{\theta}$, but only on their associated probabilities.
Here, we use the term \emph{entropy measures} in a broad definition to refer to measures of divergence
between probability distributions that belong to the family of $f$-divergencies \citep{Csis63,ali66}.
Such measures are widely used in statistics \citep{LieVaj06},
information theory \citep{CookThom91} and thermodynamics \citep{CrooksSiv11}.

The most commonly used $f-$divergence is the Kullback - Leibler (\citealp{KuLeib51})
 \begin{eqnarray}\label{KL_diver}
&& KL(p_1  \parallel  p_0)=\int_{\bs{\theta}} p_1(\bs{\theta})\log \frac{p_1(\bs{\theta})}{ p_0(\bs{\theta})}\, d\bs{\theta} \\ \nonumber
&& \hspace{2.3cm}=\int_{\bs{\theta}} p_1(\bs{\theta})\log p_1(\bs{\theta}) \,d\bs{\theta} -\int_{\bs{\theta}} p_1(\bs{\theta})\log p_0(\bs{\theta})\, d\bs{\theta} \\ \nonumber
&& \hspace{2.3cm} =-H(p_1)+cH(p_1 \parallel p_0),\nonumber
 \end{eqnarray}
with $cH(p_1 \parallel p_0)$ being the \emph{cross entropy} and $H(p_1)$ the differential entropy;
see for details in \cite{CookThom91}.
The KL-divergence is always non-negative but it is not a distance or a metric with the strict mathematical definition,
since neither the symmetry nor the triangle inequality conditions are satisfied.
In information theory, it is mostly referred to as the \emph{relative entropy} and
is a measure of the information lost when $p_0(\bs{\theta})$ is used as an approximation of $p_1(\bs{\theta})$.
Subsequently, a symmetric version of $KL$ can naturally be defined as
\begin{eqnarray}
\label{J-diver}
J(p_1, p_0)=KL(p_1  \parallel  p_0)+ KL (p_0  \parallel  p_1),\nonumber
\end{eqnarray}
\noi which dates back to Jeffreys' investigations of invariant priors (\citealp{Jeff46})
and is often called as the  \emph{ symmetrized KL-divergence} or \emph{J-divergence};
see also in \cite{Lefetal2010} for details.

The relationship between the KL-divergence and the thermodynamic integral was described
by \cite{FrielPet08} and further studied by \cite{Lefetal2010}.
In particular, the KL-divergencies between $p_1(\bs{\theta})$ and $p_0(\bs{\theta})$
can be derived  by the endpoints of the expectation of $E_{p_t}\! \big\{  U( \bs{\theta})\}$
appearing thermodynamic equation (\ref{Identity_TIgeom})  since
\begin{eqnarray}\label{KL_2edges}
	KL(p_1  \parallel  p_0)=   E_{p_1}\! \big\{  U( \bs{\theta}) \big\} -\log \lambda \mbox{~~and~~}\nonumber
  KL(p_0  \parallel  p_1)= - E_{p_0}\! \big\{ U( \bs{\theta})  \big\} +\log \lambda~.\nonumber
 \end{eqnarray}
The findings presented by \cite{Friel12} and \cite{Lefetal2010}
refer therefore to the endpoints of a geometric path.

The  question which naturally arises here is which is the role of entropy at the intermediate points for $t\in(0,1)$.
In the following, we address this issue and we illustrate how other $f-$divergencies are related to the thermodynamic integral (\ref{Identity_TIgeom}) and how can be estimated as path sampling byproducts.

\subsection{The normalised thermodynamic integral and $f-$divergencies}
\label{subsec_NTI}

In this section, we draw attention to the normalized thermodynamic integral (NTI) given by
\begin{eqnarray}\label{Identity_NTI}
NTI = \int_0^1\int_{\bs{\theta}}p_t(\bs{\theta})\,\log\frac{p_1(\bs{\theta})}{p_0(\bs{\theta})}\,d\bs{\theta}  \, dt .
\end{eqnarray}
The NTI is zero for any choices of $p_0$, $p_1$ and any geometric path $p_t$
and it can be expressed via the thermodynamic integral  using the identity
$$
NTI = \int_0^1 \int_{\bs{\theta}} p_t(\bs{\theta})\, \log \frac{q_1(\bs{\theta})}{q_0(\bs{\theta})}\,d\bs{\theta}\, dt - \log\lambda~.
$$
This identity will be used to link the thermodynamic integrals with $f-$divergencies
at any $t\in (0,1)$, generalizing the findings of \cite{Friel12} and \cite{Lefetal2010}
which associate the endpoints of the TI with KL divergencies.
To do so, we need to rewrite (\ref{Identity_NTI}) as $NTI = \int_0^1 \mathcal{KL}_t  \, dt$,
where $\mathcal{KL}_t$ is the \emph{functional KL-divergence of order $t$} defined as
\begin{eqnarray}
\label{KL_t}
&&\mathcal{KL}_t=\int_{\bs{\theta}} p_t(\bs{\theta})\log \frac{p_1(\bs{\theta})}{ p_0(\bs{\theta})}\, d\bs{\theta}
=E_{p_t} \! \big \{ U(\dn{\theta})  \big\} - \log \lambda~.
\end{eqnarray}
Then, we can express $\mathcal{KL}_t$ as the difference between the KL divergencies of $p_t$
with the two endpoint densities $p_1$ and $p_0$ since
\begin{eqnarray}
\mathcal{KL}_t&=&-cH(p_t \parallel p_1)+cH(p_t \parallel p_0)
=             KL(p_t \parallel p_1)-KL(p_t \parallel p_0).\nonumber
\end{eqnarray}
This reduces to  $\mathcal{KL}_0=-KL(p_0 \parallel p_1)$ and to
$\mathcal{KL}_1=KL(p_1 \parallel p_0)$ at the endpoints of the geometric path, which is in accordance with the findings of \cite{Friel12} and \cite{Lefetal2010}.

The divergence $\mathcal{KL}_t$ can be interpreted as a measure of \emph{relative location} of a density $p_t$
relative to $p_1$ and $p_0$.
Hence, for any $t \in [0,1]$,  $\mathcal{KL}_{t}$ indicates whether $p_{t}$ is closer to $p_0$ (negative values) or to $p_1$ (positive values).
The solution of the equation $\mathcal{KL}_{t^*}=0$ defines the point $t^*$ where
$p_{t^*}$ is equidistant (in the KL sense) from the endpoint densities.
Moreover, from (\ref{KL_t}) it is obvious that
$E_{p_{t^*}} \! \big\{ U(\dn{\theta}) \big \}$ is equal to $\log\lambda$.
Therefore, in the case that $t^*$ is known, the ratio of the normalizing constants $\lambda$
can be estimated in a single MCMC run (with $t=t^*$),
rather than employing the entire path using multiple simulations.
However this is rarely the case and, using the inverse logic, $t^*$ can be estimated by path sampling.
Having $t^*$ estimated, then the \emph{Chernoff information} can be computed in
straightforward manner (\citealp{Parzen92}, \citealp{JohSin00}, \citealp{Nielsen11}).

Following \citet{Parzen92},  the Chernoff $t$-divergence \citep{Chernoff52} is given by
\begin{eqnarray}\label{diverg_Chernoff}
C_t(p_1\parallel p_0) &=&
- \log \int_{\bs{\theta}}  p_1(\bs{\theta})^t p_0(\bs{\theta})^{1-t} d\bs{\theta}=-\log\mu(t),
\end{eqnarray}
\noi where $\mu(t)$ is the \emph{Chernoff coefficient}
\citep{Chernoff52}; also see \cite{Kakietal98} and \cite{Rauetal08}.
The key observation here is that when adopting geometric paths,
the sampling distribution $p_t(\bs{\theta})$ embodies the Chernoff coefficient since
 \begin{eqnarray}\label{p_t}
p_t(\bs{\theta}) &=&
\frac{ \big\{ z_1 p_1(\bs{\theta}) \big\} ^t \big\{ z_0 p_0(\bs{\theta}) \big\}^{1-t}}{\int_{\bs{\theta}} q_1(\bs{\theta})^t q_0(\bs{\theta})^{1-t} d\bs{\theta}}=\frac{\ p_1(\bs{\theta}) ^t  p_0(\bs{\theta}) ^{1-t}}{\mu(t)},
 \end{eqnarray}
for any $~t\in[0,1]$, which is the Boltzmann-Gibbs distribution pertaining to the Hamiltonian (energy function)
$\mathcal{H}_t(\bs{\theta})=-t\log p_1(\bs{\theta})-(1-t)\log p_0(\bs{\theta})$; see, for details, in \citet[chapter 3]{Meh10}.
In view of (\ref{p_t}) the NTI becomes
\begin{eqnarray}\label{NTIandMt}
\int_0^1\int_{\bs{\theta}}
\frac{\ p_1(\bs{\theta}) ^t  p_0(\bs{\theta}) ^{1-t}}{\mu(t)}\,
\log\frac{p_1(\bs{\theta})}{p_0(\bs{\theta})}\,d\bs{\theta}  \, dt =
 \int_0^1 \frac{d\log\mu(t)}{dt}\, dt=\bigg [ \log \mu(t)\bigg]_0^1=0,
\end{eqnarray}
since
$$
\frac{d\log\mu(t)}{dt} = \frac{1}{\mu(t)} \int
 \frac{d\{\,p_1(\bs{\theta}) ^t  p_0(\bs{\theta}) ^{1-t}\}}{dt}~ dt.
$$
From (\ref{NTIandMt}) it is straightforward to see that  the NTI  up to any point $t\in(0,1)$
is directly related to the Chernoff $t$-divergence, as described in detail in the following lemma.

\begin{lemma} \label{lemma_EntArea}
The normalised thermodynamic integral (\ref{Identity_NTI}) up to any point $t\in(0,1)$ given by
\begin{equation}
\label{NTI(t)}
NTI(t) = \int_0^t\int_{\bs{\theta}} p_t(\bs{\theta}) \log \frac{p_1(\bs{\theta})}{p_0(\bs{\theta})}\,d\bs{\theta}
\end{equation}
is equal to minus the Chernoff $t$-divergence of the endpoint densities, that is
\begin{equation}
\label{diver_Chernt}
NTI(t)= \log\mu(t)= -C_{t}(p_1\parallel p_0).
\end{equation}
\end{lemma}

    The proof of Lemma \ref{lemma_EntArea} is obtained in straightforward manner as (\ref{NTIandMt}). \hfill $\square$

Another interesting result can be obtained for $t=t^*$, the solution of the equation $\mathcal{KL}_t=0$,
and it is described in Lemma \ref{lemma_EntArea2} which follows.

\begin{lemma}
\label{lemma_EntArea2}
The Chernoff information, defined as
$$
C(p_1 \parallel p_0) = \max_{t \in [0,1]} C_t(p_1\parallel p_0)
$$
is equal to $NTI(t^*)$ with $t^*$ being the solution of equation $\mathcal{KL}_t=0$, i.e.
\begin{eqnarray}\label{diver_ChernInf}
C(p_1 \parallel p_0) = NTI (t^*) \mbox{~with~} t^* \in [0,1]: {\cal KL}_{t^*} = 0.\nonumber
\end{eqnarray}
\end{lemma}

\noi
\textbf{Proof}:
Consider the continuous and differentiable function $g(t)=NTI(t)=\log\mu(t)$.
Then $g'(t)=d\log\mu(t)/dt=\mathcal{KL}_t $
and $g''(t)=V_{p_t} \Big\{ \log\frac{p_1(\dn{\theta})}{p_0(\dn{\theta})} \Big\} > 0$;
where $V_{p_t} \Big\{ \log\frac{p_1(\dn{\theta})}{p_0(\dn{\theta})} \Big\}$
is the variance of $\log\frac{p_1(\dn{\theta})}{p_0(\dn{\theta})}$
with respect to $p_t (\dn{\theta})$.
Since $g'(t^*) = \mathcal{KL}_{t^*} =0$ and $g''(t^*)>0$,
then $g(t^*)= \min _{t \in [0,1]} \log\mu(t)$.
Hence, from (\ref{diver_Chernt}) we have that
$$
C(p_1 \parallel p_0) = \max_{t \in [0,1]} C_t(p_1\parallel p_0)
= \min_{t \in [0,1]} NTI(t) = NTI( t^* ).
$$
\hfill $\square$

The Chernoff information is often used to identify an upped bound of the probability of error
of the Bayes rule in classification problems with two possible decisions including hypothesis testing;
see \cite{Nussbaum09} and  \cite{CookThom91} for details.
It has been also used in a variety of scientific fields, primarily as a measure of similarity
between two distributions,  as for example in cryptography \citep{Baig10}.
The estimation of the Chernoff information is straightforward
and it has been treated sporadically in problem-specific cases;
see for example in \cite{Nielsen11} for computation in exponential families,
or in \cite{Julier06} for Gaussian mixture models.
The result of Lemma \ref{lemma_EntArea2} can be used to construct
a general algorithm for the estimation of the Chernoff information for any
choice of $p_1$ and $p_0$ which is described in detail in Section \ref{subsec_chernofEst}.

Before proceeding any further, we may first outline the \emph{balance property} of the NTI, which is
based on the anti-symmetry property  $C_t(p_1 \parallel p_0)=C_{1-t}(p_0 \parallel p_1)$, considered in
\cite{CrooksSiv11}.
\begin{description}
\item[The balance property:]
For any intermediate point $t\in(0,1)$ it holds that
\begin{equation}
\label{balance}
NTI(t)=- \overline{NTI}(t) \mbox{~with~}
\overline{NTI}(t)= \int_t^1\int_{\bs{\theta}} p_t(\bs{\theta}) \log \frac{p_1(\bs{\theta})}{p_0(\bs{\theta})}\,d\bs{\theta}
\end{equation}
and therefore the maximum absolute value occurs at $t^*$ and it is equal to NTI$(t^*)$.
\end{description}

\noi Based on Lemmas \ref{lemma_EntArea} and \ref{lemma_EntArea2} and the balance property,
it occurs that the Chernoff $t-$divergences  (either from  $p_1$ to $p_0$ or in the opposite direction)
can be directly computed from the NTI.
Subsequently, a number of other divergencies related to  Chernoff can be obtained from NTI.
The \emph{Bhattacharyya distance} \citep{Bhatt43} occurs at $t=0.5$, that is
\begin{eqnarray}\label{diverg_Bah}
&& Bh(p_1,p_0)= C_{0.5}(p_1\parallel p_0)=-\log \int_{\bs{\theta}} \sqrt{p_1(\bs{\theta}) p_0(\bs{\theta})} d\bs{\theta}=-\log \rho_B.\nonumber
 \end{eqnarray}
\noi  The Bhattacharyya coefficient $\rho_B$  can be implemented in turn to derive the  \emph{Bhattacharyya–-Hellinger distance}
 \citep{Bhatt43,Helling09} since
$He(p_1,p_0)= \sqrt{1-\rho_B}$.
Based on the Chernoff $t$-divergence we may also derive the \emph{R\'{e}nyi $t$-divergence}
$R_t(p_1\parallel p_0)
= C_{t}(p_1\parallel p_0)/(1-t)$ \citep{Renyi61}
and the Tsallis $t$-relative entropy
$T_t(p_1\parallel p_0)
= \big[\exp\big\{-C_{t}(p_1\parallel p_0)\big\}-1\big]/(1-t)$.

\begin{figure}[h!]
\begin{center}
  \includegraphics[width=13cm]{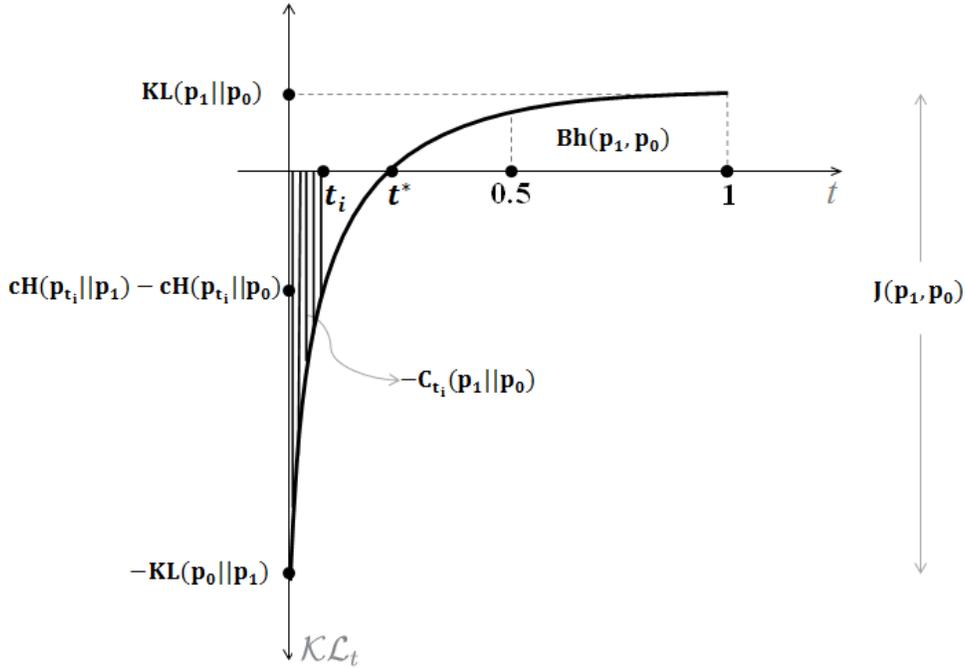}\\
  \caption{\scriptsize Graphical representation of the NTI: the plot of $\mathcal{KL}_t(\bs{\theta})$ over $t$.}
   \label{fig_NTI}
   \end{center}
\end{figure}

A graphical representation of the NTI is given in Figure \ref{fig_NTI}.
The cross entropy differences between $p_t$
and the endpoint distributions ($p_0$ and $p_1$)
are depicted on the vertical axis.
The KL-divergencies between $p_0$ and $p_1$ are located at the endpoints of $[0,1]$.
Their difference represents the $J-$divergence.
From Lemma \ref{lemma_EntArea}, the Chernoff $t-$divergence for any $t_i \in [0,1]$
is given by the area between the curve and the $t$-axis
from $t=0$ to $t=t_i$.
The Chernoff information is given by the corresponding area up to $t=t^*$
while the Bhattacharyya distance is given by the corresponding area from zero up to  $t=0.5$.

To sum up, in this section we illustrated how entropy measures are directly associated
with the NTI.
For this reason, all these measures  can be derived using path sampling.
Hence, the NTI given in (\ref{Identity_NTI}) can offer another link between
Bayesian inference, information theory and thermodynamics (or statistical mechanics).
For instance, under the Hamiltonian $\mathcal{H}_t(\bs{\theta})$, \citet[Section 3.3]{Meh10}
discuss the \emph{excess} or \emph{dissipated} work in thermodynamics and
its relation to the data processing theorem in information theory, with the NTI emerging
in the case of reversible processes. In a more general framework, \cite{CrooksSiv11}
consider \emph{conjugate trajectories}, that is forward (from $t=0$ to $t=1$)
and backward processes (from $t=1$ to $t=0$), to derive the physical significance of the
$f-$divergencies considered here, in terms of non-equilibrium dynamics.
Note also that the balance property (\ref{balance}) satisfies the (recently derived) equality of \cite{Jarzynski97} \
and confirms \cites{Crooks99} theorem; see, for details, in \cite{Meh10} and \cite{CrooksSiv11}.
Further parallelism between the NTI and statistical mechanics  is not attempted here, leaving this part to the experts on the field.
In the next section we focus on the study of the MCMC estimators of  $\log \lambda$ constructed using TI and geometric paths.
We further study and analyse how the $f-$divergencies can be estimated as path sampling byproducts.

\subsection{MCMC path sampling estimators}
\label{sec_estimators}

Numerical approaches are typically used to compute the external integral of (\ref{Identity_TI}), such as the trapezoidal or Simpson's rule (\citealp{Ogata89}; \citealp{Neal93}; \citealp{GelMeng98}, among others).
The numerical approaches require the formulation of an  $n$-point discretisation
${\cal T} = \{t_0, t_1, \dots, t_n\}$ of $[0,1]$, such that  $0=t_0 < ... < t_{n-1}<t_n=1$,
which is called \emph{temperature schedule}.
A separate MCMC
run is performed at each $t_i$ with target distribution the corresponding $p(\bs{\theta}|\,t_i)$, $i=0,...,n$.
The MCMC output is then used to estimate  ${\cal E}_{t} = E_{p_{t}}\!\!\left\{ U(\bs{\theta})\right\}$
by the sample mean $\widehat{\cal E}_t$  of the simulated values $\{\bs{\theta}^{(r)}\}^{R}_{r=1}$ generated
from $p_{t}$ for each $t\in {\cal T}$. The final estimator
is derived by
\begin{eqnarray}\label{estimator_TI}
 \log\widehat{\lambda} &=& \sum_{i=0}^{n-1}(t_{i+1}-t_i)
  \frac{\widehat{\mathcal{E}}_{t_{i+1}} + \widehat{\mathcal{E}}_{t_{i}}}{2};
 \end{eqnarray}
see also in \cite{FrielPet08}.

At a second step, the posterior output at each $t_i$ and $\log\widehat{\lambda}$
can be employed to estimate $t^*$ and the Chernoff information.
Here we provide an algorithm for that purpose, which yields also the estimated Chernoff $t-$divergencies
for  any $t\in(0,1)$ and subsequently the $f-$divergencies described in Section \ref{subsec_NTI}.

\subsubsection{Estimation of the Chernoff $t-$ divergencies and information}
\label{subsec_chernofEst}

Estimating the Chernoff information is generally a non-trivial and cumbersome procedure.
For instance, \cite{Nielsen11} describe a  \emph{geodesic bisection optimization algorithm}
that approximates  $C(p_1 \parallel p_0)$ for multidimensional distributions
which belong to the exponential family, based on Bregman divergences (named after Bregman, who introduced the concept
in \citealp{Breg67}). \cite{Julier06} provides also an approximation for Gaussian mixture models.
Here we introduce a TI based MCMC method for the estimation of Chernoff information which can
be used for any choice of $p_0$ and $p_1$ distributions.

Following Lemma \ref{lemma_EntArea2}, the Chernoff information is given by $NTI(t^*)$.
Therefore, in order to compute the Chernoff information we need first to estimate $t^*$ for which  $\mathcal{KL}_{t^*}$ is zero.
The computation  of $t^*$ can be achieved by adding a
number of steps in the path sampling procedure according to the following algorithm.

\begin{description}
\label{ChernoffAlgo}
    \item[Step 1]   Perform $n$ MCMC runs to obtain $\widehat{\cal E}_t$ for all $t \in {\cal T}$
                    and $\log\widehat{\lambda}$ from (\ref{estimator_TI}).
    \item[Step 2]   Calculate $\widehat{\mathcal{KL}}_t = \widehat{\mathcal{E}}_t - \log\widehat{\lambda}$ for all $t \in {\cal T}$.

    \item[Step 3]   Identify interval $\left(t_{i^*}^-, t_{i^*+1}^+\right)$ where the sign of $\mathcal{KL}_t$ changes;
    								where
    								$$
    								t_i^- = \max \! \big( t \in {\cal T} : \widehat{\mathcal{KL}}_t < 0 \big) \mbox{~~and~~}
    								t_i^+ = \min  \! \big( t \in {\cal T} : \widehat{\mathcal{KL}}_t > 0 \big)~.
    								$$
    								Note, that $\mathcal{KL}_t$ will be negative for any $t<t^*$ and positive otherwise since
    								since $\frac{d {\cal KL}_t }{dt} = V_{p_t} \Big\{ \log \frac{p_1(\dn{\theta})}{p_0(\dn{\theta})} \Big\}>0$
    								and therefore ${\cal KL}_t$ it is an increasing function of $t$.

    \item[Step 4]   Perform extra MCMC cycles by further discretising $\left(t_{i^*}^-, t_{i^*+1}^+\right)$
    								until the required precision is achieved.
    								
    \item[Step 5]   Update ${\cal T}$ and $n$ to account for the new points $t_i\in \left(t_{i^*}^-, t_{i^*+1}^+\right)$ used in Step 5.
    								
    \item[Step 6]   Once the $t^*$ is estimated,
    				the MCMC output already available from the runs in Steps 1 and 4 can be used to estimate  the Chernoff information.
    				In particular, it is estimated as described in (\ref{estimator_TI})
    				having substituted $\widehat{\mathcal{E}}_t$ by $\widehat{\mathcal{KL}}_t$ for all $t\in{\cal T}$
    				and only accounting for $t_i \leq t^*$ in the summation.
    				Therefore, the Chernoff information is estimated by $\widehat{NTI}(t^*)$ given by
										\begin{eqnarray}										
										\log\widehat{NTI}(t^*) &=& \sum_{ i \in {\cal I}:\, t_{i+1} \leq \,t^* } (t_{i+1}-t_i)
										                           \frac{\widehat{\mathcal{KL}}_{t_{i+1}} + \widehat{\mathcal{KL}}_{t_{i}}}{2} \nonumber\\
										                       &=& \sum_{ i \in {\cal I}:\, t_{i+1} \leq\, t^* } (t_{i+1}-t_i)
										                           \frac{\widehat{\mathcal{E}}_{t_{i+1}} + \widehat{\mathcal{E}}_{t_{i}} }{2} - t^* \log \widehat{\lambda}~ ,
										                           \label{estimator_NTI}
										 \end{eqnarray}
    								where the ${\cal I} = \{ 0, 1, \dots, n\}$ and $n=|{\cal T}|$.
\end{description}
In the special case where the path sampling is combined with output from MCMC algorithms which involve
tempered transitions (see \citealp{CalGir09} for details), the estimation of the Chernoff information
comes with low computational cost.
This approach can be attractive and useful in the case of multi-modal densities.
The same algorithm can be also implemented to compute the rest of the f-divergencies measures discussed in Section \ref{subsec_NTI}.
In fact, their estimation is less demanding since it requires one additional MCMC run, in order to derive
the estimated $\mathcal{KL}_{t_i}$ at the point of interest;
for instance at $t_i$=0.5 we derive the $Bh(p_1,\,p_0)$ and $He(p_1,\, p_0)$ divergencies.

\subsection{Error, temperature schedule and geometric perspective}
\label{sub_ErrorSched}

In this section we study two important sources of error for path sampling estimators:
the {\it path-related variance} and the {\it discretisation error}.
The path-related variance is the error related to the choice of the path which, for geometric ones,
is restricted to the selection of the endpoint densities.
On the other hand, for any given path, the discretisation error is related to the choice of the temperature schedule ${\cal T}$
and is derived from the numerical approximation of the integral over $[0,1]$.
In order to examine these two error sources,
we provide a geometric representation of TI (eq. \ref{Identity_TIgeom}) and NTI (eq. \ref{Identity_NTI}) identities.
This leads us to a better understanding of the behaviour of the path sampling estimators.

\subsubsection{Path-related variance}
\label{pathvariance}

The total variance of $\log \widehat{\lambda}$ has been reported by \cite{GelMeng98}
in the case of stochastic $t$ with an appropriate prior distribution attached to it.
Further results were also presented by \cite{Lefetal2010} for geometric paths.
They have showed that the total variance is associated with the $J-$divergence of
the endpoint densities and therefore with the choice of the path.
Here we focus on the $t$-specific variances $V_t=V_{p_t}\{U(\bs{\theta})\}> 0$ of $U(\bs{\theta})$
(hereafter \emph{local variance}) which are the components of the total variance.

\begin{figure}[h!]
\begin{center}
  \includegraphics[width=11cm]{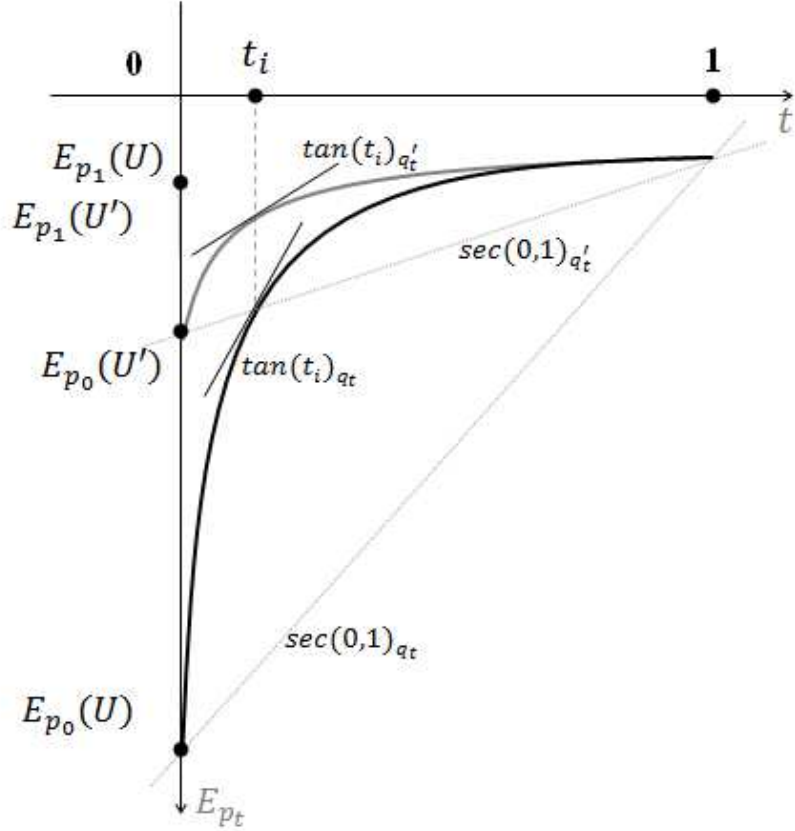}\\
  \caption{\footnotesize{Graphical representation of the TI: the plot of the curve ${\cal E}_t = E_{p_t}\{U( \bs{\theta})\}$ over  $t$,
                         based on two paths $q_t$ (black line) and $q'_t$ (grey line). For
                         each path, the $J-$distance between the endpoints coincides with the slope of the corresponding
                        secant, $sec(0,1)$. The slope of the tangent $tan(t_i)$ equals the local variance $V_{t_i}$.}}
                        \label{fig_2paths}
            \end{center}
\end{figure}

Figure \ref{fig_2paths} is a graphical representation of TI.
To be more specific, the curve represents the ${\cal E}_t $ values for each $t\in[0,1]$
while the area between the t-axis and the curve gives the thermodynamic integral (\ref{Identity_TI}).
In this figure, the error of the TI estimators is depicted by the steepness of the curve of ${\cal E}_t$.
This result is based on the fact that the \emph{partition function} $z_t$ is the cumulant
generating function of $U(\bs{\theta})$  \citep[section 2.4]{Meh10}
and therefore the first derivative of ${\cal E}_t$ is given by the local variance $V_t$,
that is ${\cal E}_t'  = V_t$. It follows that the slope of the tangent of the curve at each $t$ equals  to $V_t$.
Therefore, the graphical representation of two competing paths can provide valuable information
about the associated variances of their corresponding estimators.

In the case of geometric paths particularly, $J(p_1,\,p_0)$ coincides with the slope of the
secant defined at the endpoints of the curve and lays below the curve of the strictly increasing (in terms of $t$) function
${\cal E}_t$.
Therefore, it can be used as an indicator of the slope of the curve and the
result of \cite{Lefetal2010} has a direct visual realisation.
The result can be generalised for any other pair of successive points,
say $(t_i, {\cal E}_{t_i})$ and $(t_{i+1}, {\cal E}_{t_{i+1}})$,
with the corresponding slope (or gradient) of the secant $sec(t_i,t_{i+1})$ given by
 \begin{eqnarray}\label{secant}
\nabla sec(t_i,t_{i+1})
&=&\frac{ {\cal E}_{t_{i+1}} - {\cal E}_{t_{i+1}} }{t_{i+1}-t_i}
 = \frac{ {\cal KL}_{t_{i+1}} -{\cal KL}_{t_{i}}}{t_{i+1}-t_i}.
\end{eqnarray}
\noi The latter  is derived from (\ref{KL_t}) and it reflects the fact that
 the slopes of the curves depicted in Figures \ref{fig_NTI} and \ref{fig_2paths} are identical.
Additionally, ${\cal KL}_t$ can be written in terms of the KL-divergence between the
successive sampling densities $p_{t_i}$ and $p_{t_{i+1}}$ since,  from (\ref{p_t}) we obtain
\begin{eqnarray}
\label{klt_andKL}
KL(p_{t_i} \parallel p_{t_{i+1}})
& = & \int_{\boldsymbol{\theta}} p_{t_i}(\bs{\theta})\log \big \{p_1(\bs{\theta})^{t_i-t_{i+1}}p_0(\bs{\theta})^{t_{i+1}-t_{i}}\big\} \,d\boldsymbol{\theta}+\log\frac{\mu(t_{i+1})}{\mu(t_{i})}\nonumber\\
& = & -(t_{i+1}-t_i) {\cal KL}_{t_i}+\log\frac{\mu(t_{i+1})}{\mu(t_{i})}.
\end{eqnarray}

\noi Using (\ref{secant}) and (\ref{klt_andKL}), we can associate the $J-$divergence between two successive points
with the slope of the secant $sec(t_i,t_{i+1})$  since
\begin{eqnarray}\label{JandSec}
\nabla sec(t_i,t_{i+1})=\frac{J(p_{t_{i}},p_{t_{i+1}})}{(t_{i+1}-t_i)^2}
\end{eqnarray}
generalizing the result of \cite{Lefetal2010} for the endpoints of the graph
where the slope of the $sec(0,1)$ is given by $J(p_1,p_0)$.
For successive points closely placed to each other (that is, for $\Delta(t_{i})=t_{i+1}-t_{i} \rightarrow 0$)
the slope of the secant approximates the corresponding slope of the tangent of the curve and therefore the local variance.
Hence, the $J-$divergence between any two successive points is indicative of the slope of the
curve and consequently of the  associated variance.
For example, in Figure \ref{fig_2paths} for values of $t$ close to zero
 the slope of curve  is very steep indicating high local variability.


The local variances of the path sampling estimators discussed here depend on the selection of the path.
In the next section,
we proceed with the study of the discretisation error and its effect on the path sampling estimators based
on both the TI  and NTI  identities for any fixed geometric path.

\subsubsection{Discretisation error}

\cite{CalGir09} expressed the discretisation error in terms of differences of relative entropies of
successive (in terms of $t$) sampling distributions.
The result of \cite{CalGir09} can be written for any geometric path as follows
 \begin{eqnarray}\label{CG_bias}
\log \widehat{\lambda}  = \sum_{i=0}^{n-1} \frac{\widehat{z}_{t_{i+1}}}{\widehat{z}_{t_{i}}}&=&
 \frac{1}{2}\sum_{i=0}^{n-1}  \, (t_{i+1}-t_i) \, \left\{ \widehat{\mathcal{E}}_{t_{i+1}}+\widehat{\mathcal{E}}_{t_{i}} \right \} \\
&&  +~\frac{1}{2} \sum_{i=0}^{n-1}  \left\{\widehat{KL}(p_{t_i}\parallel \,p_{t_{i+1}})-\widehat{KL}(p_{t_{i+1}}\parallel p_{t_{i}}) \right\}, \nonumber
 \end{eqnarray}
\cite{CalGir09} consider the case for $\Delta(t_i) \rightarrow 0$ in (\ref{CG_bias})
and outline that the first summation is equivalent to the trapezium rule used for numerical integration
with the associated error expressed in terms of the asymmetries between the KL divergencies defined between $p_{t_i}$ and $p_{t_{i+1}}$.
In view of (\ref{klt_andKL}), expression \ref{CG_bias} becomes
 \begin{eqnarray}\label{NTIdis_error}
\log \widehat{\lambda} &=& \frac{1}{2}\sum_{i=0}^{n-1}  \, \Delta(t_i) \,   \left\{\widehat{\mathcal{E}}_{t_{i+1}}+\widehat{\mathcal{E}}_{t_{i}} \right \}
-~\frac{1}{2} \sum_{i=0}^{n-1} \, \Delta(t_i) (\widehat{\mathcal{KL}}_{t_i}+\widehat{\mathcal{KL}}_{t_{i+1}}),
 \end{eqnarray}
\noi since $ \sum_{i=0}^{n-1}\log \frac{\mu(t_i)}{\mu(t_{i+1})}=0$.
The second term in the left side of (\ref{NTIdis_error}) is the approximation of the NTI (using the trapezoidal rule), which
indeed it should be zero.
According to the discussion in Section \ref{pathvariance},
the relative entropies in (\ref{CG_bias}), as well as the
areas above and below the $t$-axis which represent the Chernoff divergencies, are not expected to be zero.
They both represent  the path-related variance which is independent (and pre-existing) of the discretisation error.
The discretisation error consists of the asymmetries that occur under any particular tempering schedule
either in the TI or in NTI. The symmetry is a feature of the thermodynamic integration
and it represents the trade-off between uncertainty in the forward and backward trajectories.
Therefore, the error manifests as lack of symmetry in the assessment of the uncertainty
due to the discretisation, as explained below.

While the path-related variance is independent from the discretisation error, the
reverse argument does not hold. In fact, the discretisation error is highly influenced and dependent upon the
path-related variance. Consider two pairs of successive points, located close to the zero and unit endpoints
in Figure \ref{fig_NTI}, say $t^{(0)}_i, t^{(0)}_{i+1}$  and $t^{(1)}_j, t^{(1)}_{j+1}$ respectively, for $i,j=1,...,n$.
Further assume that the distances between the points within each pair are equal, say $\delta > 0$.
For the first pair, the corresponding $\mathcal{KL}_t$s on the vertical axis are distant due to the steepness of the curve.
On the contrary, for the second pair the corresponding $\mathcal{KL}_t$s are very close,
due to the fact that the slope of the curve is almost horizontal. Therefore, using the trapezoidal rule,
for equally spaced pairs of points we approximate a large part of the curve towards the zero end
and a small part of the curve towards the unit end. In order to achieve the same degree of accuracy at both ends,
the second pair of points need to be closer.
In conclusion, the temperature schedule should place more points
towards the end of the path where the uncertainty (slope) is higher.
For instance, the powered fraction (PF) schedule \citep{FrielPet08}
\begin{equation}\label{schPrior}
\mathcal{T}_{PF}=\{t_i\}_{i=1}^n   \mbox{~~such as~~} t_i=(1/n)^\mathcal{C}, \mathcal{C}=1/a>1,
 \end{equation}
places more points towards the zero endpoint of the path. \cite{Xieetal11}
proposed a closely related geometric schedule where the $t_i$s are chosen according to
evenly spaced quartiles of a  $Beta(a,1)$ distribution.
Recently, \cite{Friel12} proposed an
adaptive algorithm for the temperature schedule that takes under consideration the local variances
in order to locate the high uncertainty points. The algorithm traces the points on
the curve and assigns more $t_i$s close to their regions. The gain in the error is then achieved with
a small computational price.

\section{Bayesian model comparison using tempered transitions}
\label{sec_Estimators}

Let us consider two competing models, $m_1$ and $m_0$, with equal prior probabilities.
Then, the Bayes factor (BF; \citealp{Jeff35,Jef61,kas:raf95})
is derived as the ratio of the marginal likelihoods
\begin{equation}
\label{marginal}
f(\dn{y}|\,\mii) = \int_{\bs{\theta}} f (\dn{y} |\,\boldsymbol{\theta},\mii) \pi(\boldsymbol{\theta}|\,\mii)\, d\boldsymbol{\theta} \end{equation}
for each model $m_1$ and $m_0$;
where $\dn{y}$ denotes the data matrix and $\pi(\boldsymbol{\theta}|\mii)$ is the prior density of the parameter vector
under the model $m_i$.
The integral involved in the marginal likelihood (eq. \ref{marginal})
is often high dimensional  making its analytic computation infeasible.
Therefore a wide variety of MCMC based methods have been developed for its estimation;
see , for example, in \cite{chib95}; \cite{GelMeng98}; \cite{LewRaf97} among others.

Since the marginal likelihood is simple the normalizing constant of the posterior distribution $f(\boldsymbol{\theta}|\dn{y},\mii)$
and can be estimated by path sampling.
Recently, such methods have been considered for marginal likelihood estimation by \cite{LarPhil06}, \cite{FrielPet08} and \cite{Lefetal2010}.

\subsection{The stepping-stone identity}
In this section we consider an alternative approach that is based on the stepping-stone sampling, presented by
\cite{Xieetal11} and \cite{Fanetal11} for the estimation of the marginal likelihood. Closely related ideas are also discussed
in the context of the free energy estimation in \citet[see section 6.2 and references within]{Neal93}.
The stepping-stone sampling considers finite values $t_i \in {\cal T}$, that are placed according to a temperature schedule
as the ones discussed in Section \ref{sub_ErrorSched}. The ratio of the normalizing constants can be expressed
as
\begin{equation} \label{def_ratioSS}
\lambda=
\frac{z_{1}}{z_{0}}=\frac{z_{t_n}}{z_{t_{n-1}}}\frac{z_{t_{n-1}}}{z_{t_{n-2}}} \dots \frac{z_{t_{1}}}{z_{t_0}}=\prod_{i=0}^{n-1} \frac{z_{t_{i+1}}}{z_{t_{i}}}.\nonumber
\end{equation}
\noi Hence, the ratio of the normalizing constants can be estimated using $z_{t_{i+1}}/z_{t_{i}}$  as an  intermediate step
 that can be estimated from $t$ specific MCMC samples  based on the identity
\begin{eqnarray}\label{identity_SSeach}
&& \frac{z_{t_{i+1}}}{z_{t_{i}}}= \int_{\bs{\theta}} \frac{q_{t_{i+1}}(\bs{\theta})}{q_{t_{i}}(\bs{\theta})} ~p_{\,t_{i}}(\bs{\theta}) \,d\bs{\theta}; \nonumber
\end{eqnarray}
see \cite{Xieetal11} for details. For geometric paths, the stepping-stone identity for $\lambda$ is then given by
\begin{eqnarray}\label{Identity_SS}
 \lambda= \prod_{i=0}^{n-1}\int_{\bs{\theta}}\left\{\frac{q_1(\bs{\theta})}{q_0(\bs{\theta})}\right\}^{\Delta(t_i)}p_{\,t_{i}}(\bs{\theta})\,d\bs{\theta}.
\end{eqnarray}
\cite{Xieetal11} presented the stepping-stone sampling specifically
for estimating the marginal likelihood (under a certain geometric path)
while \cite{Fanetal11} modified the initial marginal likelihood estimator
in order to improve its properties (both estimators are addressed later on in this section).
However, as outlined here, the stepping-stone sampling can be considered as a general method,
alternative to path sampling, that can be applied for the estimation of ratios of unknown normalized constants.

Hence, identities (\ref{Identity_TIgeom}) and  (\ref{Identity_SS}), are two closely related alternative
tempered transition methods for the estimation of normalizing constants using geometric paths.
Any estimator developed via thermodynamic integration has its corresponding stepping-stone estimator  and vise versa. \
In the next section, we present existing methods classified by the tempered method that has been originated and the adopted path.
This method-path approach allows us to further introduce new estimators based on the counterpart existing ones.

\subsection{Marginal likelihood estimators}
\label{sec_marg}

In order to avoid confusion, hereafter we will name each estimator based
on the method (thermodynamic or stepping-stone) and on the path implemented for its derivation.

The  power posteriors (\citealp{LarPhil06}, \citealp{FrielPet08})  and the
the stepping stone \citep{Xieetal11} marginal likelihood estimators
are  using the same geometric path but they are based on different identities,
approaching the same problem using a different perspective.
Both methods implement the geometric \emph{ prior-posterior} path, namely
 \begin{eqnarray}
 \label{Path_PP}
q_t^{\emph{PP}}(\bs{\theta})=\left \{f(\dn{y}|\,\bs{\theta})\pi(\bs{\theta})\right \}^t\pi(\bs{\theta})^{1-t}=f(\dn{y}|\,\bs{\theta})^t\pi(\bs{\theta}),
\end{eqnarray}
where  $q_0(\bs{\theta})=\pi(\bs{\theta})$ is a proper prior for the model parameters
 and  $q_1(\bs{\theta})=f(\bs{\theta|}\,\dn{y})\,\pi(\bs{\theta})$  is  the corresponding unnormalized posterior density.
Setting the prior-posterior in (\ref{Identity_TIgeom}) and (\ref{Identity_SS}), yields  the thermodynamic and the stepping-stone
prior-posterior identities  (PP$_T$ and PP$_S$ respectively) for the marginal likelihood
\begin{eqnarray}\label{identities_PP}
\log f(\dn{y}) =\int_0^1 E_{p_t^{\emph{PP}}}\left \{\log f(\dn{y}|\,\bs{\theta})\right \}\,dt \mbox{~~and~~}
f(\dn{y})= \prod_{i=0}^{n-1}\int_{\bs{\theta}}\left\{ \log f(\dn{y}|\,\bs{\theta}) \right\}^{\Delta(t_i)}p_{\,t_{i}}^\emph{PP}(\bs{\theta})\,d\bs{\theta}\nonumber
\end{eqnarray}
where $p_t^\emph{PP}(\bs{\theta}|\dn{y})$ is the density normalized version of (\ref{Path_PP}).

\cite{Fanetal11} modified the estimator of \cite{Xieetal11} using instead the \emph{importance-posterior} path
\begin{eqnarray}\label{Path_IPP}
q_t^\emph{IP}(\bs{\theta})=\left\{f(\dn{y}|\,\bs{\theta})\,\pi(\bs{\theta})\right\}^tg(\bs{\theta})^{1-t}.\nonumber
\end{eqnarray}
\noi
The importance posterior path was one of the paths that \cite{Lefetal2010} considered
for the estimation of the marginal likelihood. It should be noted that the
density $g(\bs{\theta})$ is required to be proper so that $z_0=1$.
It can be constructed by implementing the posterior moments available
from the MCMC output at $t=1$. The thermodynamic and stepping-stone importance-posteriors
(IP$_T$ and IP$_S$ respectively) are derived by the identities
\begin{eqnarray}
\log f(\dn{y})  &=& \int_0^1 \!\! E_{p_t}^\emph{IP}\left[\log  \frac{f(\dn{y}|\,\bs{\theta})\,\pi(\bs{\theta})}{g(\bs{\theta})}\right]dt  \mbox{~~and~~} \label{identities_IP_t} \\
f(\dn{y}) & =  & \prod_{i=0}^{n-1}\int_{\bs{\theta}}\left\{ \frac{f(\dn{y}|\,\bs{\theta})\,\pi(\bs{\theta})}{g(\bs{\theta})} \right\}^{\Delta(t_i)}
\!\!\! p_{\,t_{i}}^\emph{IP}(\bs{\theta})\,d\bs{\theta},  \nonumber 
\end{eqnarray}
\normalsize
where $p_t^\emph{IP}(\bs{\theta})$ is the density normalized version of $q_t^\emph{IP}(\bs{\theta})$.

The TI identity  appearing in (\ref{identities_IP_t}) 
has the attractive feature of sampling from $g(\bs{\theta})$, rather than the prior,
for $t=0$. It also retains the stability ensured by averaging in log scale according to the thermodynamic approach.
Therefore, in specific model settings, the estimators based on the thermodynamic importance posteriors
can perform more efficiently than estimators based on the other expressions,
provided that an importance function can be formulated. It is our belief that beyond the four expressions reviewed here,
others may be developed within this broad framework, by choosing the appropriate path for particular models,
coming with thermodynamic and stepping-stone variants.

\subsection{Bayes factor direct estimators}\label{sec_BF}

The BF is  by definition a ratio of normalized constants. Therefore, (\ref{Identity_TIgeom}) and  (\ref{Identity_SS})
can be implemented to construct direct BF estimators, rather than applying the methods to each model separately.
\cite{LarPhil06} implemented the thermodynamic integration, in order to link two competing (not necessary nested) models, instead of densities.
That was achieved by  choosing the appropriate path, in a way that eventually produces directly a BF estimator.
\cite{LarPhil06} were motivated by the fact that lack of precision on each  marginal likelihood  estimation, may alter the BF
interpretation. They argue, that a simultaneous estimation of the two constants can ameliorate that to some extend. The idea is to
employ a bidirectional \emph{melting-annealing} sampling scheme, based on the \emph{model-switch} path:
\begin{eqnarray}\label{Path_MS}
 q_t^{\emph{MS}}(\bs{\theta})=\left\{f(\dn{y}|\,\bs{\theta},\mi)\,\pi(\bs{\theta}| \, \mi)\right\}^{t}\left\{f(\dn{y}|\,\bs{\theta},\mo)\,\pi(\bs{\theta}| \, \mo)\right\}^{1-t}.\nonumber
\end{eqnarray}
\noi \cites{LarPhil06} thermodynamic model-switch  (MS$_T$) identity for the BF and its stepping-stone counterpart (MS$_S$)
are as follows
\begin{eqnarray}\label{identities_QMS}
&&\log BF_{10}=\int_0^1 E_{p_t^{\emph{MS}}}\left[\log \left\{\frac{f(\dn{y}|\,\bs{\theta},\mi)\,\pi(\bs{\theta}| \, \mi)}{f(\dn{y}|\,\bs{\theta},\mo)\,\pi(\bs{\theta}| \, \mo)}\right\}  \right]dt  \nonumber\\
\mbox{and} \nonumber\\
&&BF_{10}= \prod_{i=0}^{n-1}\int_{\bs{\theta}}\left\{\frac{f(\dn{y}|\,\bs{\theta},\mi)\,\pi(\bs{\theta}| \, \mi)}{f(\dn{y}|\,\bs{\theta},\mo)\,\pi(\bs{\theta}| \, \mo)}\right\}^{\Delta(t_i)}p_{\,t_{i}}^{\emph{MS}}(\bs{\theta}|\,\dn{y})\,d\bs{\theta},\nonumber
\end{eqnarray}
\noi  where the expectation is taken over $p_t^{\emph{MS}}(\bs{\theta}|\dn{y})$ which is the density
obtained after normalizing the model-switch path $q_t^{\emph{MS}}(\bs{\theta})$.
In case where $\boldsymbol{\theta}$ is common between the two models (for instance
if the method is used to compare paths under different endpoints, see \citealp{LarPhil06} for an example)
the method is directly applicable.
Otherwise, if  $\bs{\theta}=( \bs{\theta}_{m_1},\bs{\theta}_{m_0} )$,  pseudo-priors need to be assigned at the
endpoints of the path.

Having in mind the direct estimation of Bayes factors,
more complicated estimators may be derived using \emph{compound} geometric paths.
With the term compound paths we refer to paths that consist of
a \emph{hyper} geometric path,
$Q_t (\dn{\theta}) = Q_1(\dn{\theta})^t Q_0(\dn{\theta})^{1-t}$,
used to link two competing models
and a \emph{nested} path $q_t(\dn{\theta},i)$ for each endpoint function $Q_i$, for $i=0,1$.
The two intersecting paths form a \emph{quadrivial}, $(Q\circ q)_t(\bs{\theta})$ with $t\in[0,1]$
that can be defined as
\begin{eqnarray}\label{identities_QuadMulti}
&&(Q\circ q)_t(\bs{\theta})=
\big[ q_1(\bs{\theta},1)^t q_0(\bs{\theta},1)^{1-t}\big]^t\,
\big[ q_1(\bs{\theta},0)^t q_0(\bs{\theta},0)^{1-t}\big]^{1-t}.\nonumber
\end{eqnarray}
The multivariate extension is discussed in detail in \cite{GelMeng98}.
The endpoint target densities are given by
$q_i(\bs{\theta},i)$ for $t=0$ and $t=1$ respectively estimating the ratio
$z_{1}/z_{0} = \int q_1(\bs{\theta},1) d\dn{\theta} \times \left[ \int q_0(\bs{\theta},0) d\dn{\theta} \right]^{-1}. $
The densities $q_i(\bs{\theta},j)$ for $i,j = 0, 1$ and $i \neq j$  serve as linking densities within each nested path.
Therefore, following the importance-sampling logic, they should play the role of  approximating (importance)
functions for each $q_i(\bs{\theta},i)$.

For the specific case of the Bayes factor estimation,
the objective is to retrieve the marginal likelihoods at the endpoints
and therefore it is reasonable to consider as nested paths the prior-posterior and the importance-posterior paths,
discussed in the previous section. The importance-posterior BF quadrivial, for instance, is as follows
\begin{eqnarray*}\label{Path_geomrefBF}
(Q\circ q)^{{IP}}_t(\bs{\theta}) &=&
\Big[\big\{f(\dn{y}|\,\bs{\theta},\mi) \pi(\bs{\theta}|\,\mi)\big\}^{t}  g(\bs{\theta}|\,\mi)^{1-t}\Big]^{t}  \\
&& \hspace{4cm}  \times
\Big[\big\{f(\dn{y}|\,\bs{\theta},\mo) \pi(\bs{\theta}|\,\mo)\big\}^{1-t}g(\bs{\theta}|\,\mo)^{t} \Big]^{1-t}\nonumber
\end{eqnarray*}
 \noi leading to the thermodynamic ($Q_{IP_T}$) and stepping-stone ($Q_{IP_S}$) expressions
\begin{eqnarray}\label{Identity_QIPP}
&&\log BF_{10} =\int_0^1 E_{P_t}\left[ \log \frac{\big\{f(\dn{y}|\,\bs{\theta},\mi)\,\pi(\bs{\theta}|\,\mi)/g(\bs{\theta}|\,\mi)\big\}^{2t}g(\bs{\theta}|\,\mi)} {\big\{f(\dn{y}|\,\bs{\theta},\mo)\,\pi(\bs{\theta}|\,\mo)/g(\bs{\theta}|\,\mo)\big\}^{2(1-t)}g(\bs{\theta}|\,\mo)}\right]dt \nonumber \\
\mbox{and}\nonumber \\
&& BF_{10}=\prod_{i=0}^{n-1}\int_{\bs{\theta}} \log \frac{\big\{f(\dn{y}|\,\bs{\theta},\mi)\,\pi(\bs{\theta}|\,\mi)/g(\bs{\theta}|\,\mi)\big\}^{2 T_i }g(\bs{\theta}|\,\mi)} {\big\{f(\dn{y}|\,\bs{\theta},\mo)\,\pi(\bs{\theta}|\,\mo)/g(\bs{\theta}|\,\mo)\big\}^{2(1-T_i)}g(\bs{\theta}|\,\mo)} \,P_{t_i}(\bs{\theta}) \, d \bs{\theta}, \nonumber
\end{eqnarray}

\noi where $P_t(\bs{\theta})=(Q\circ q)^{{IP}}_t (\bs{\theta})=/Z_t$, $Z_t= \int_{\bs{\theta}}(Q\circ q)^{{IP}}_t \,d\bs{\theta}$, $t\in[0,1]$.
In the thermodynamic expression, $t$ is the \emph{melting} temperature and $1-t$  the \emph{annealing} one,
assuming that the procedure starts at $t=0$ and gradually increases  to $t=1$.
The hyper-path ensures that while the model $m_1$ is melting, the model  $m_0$  is annealing. At the same time, the importance-posterior path serving
as the nested one, links the posterior with the importance at each model separately. In  the stepping-stone counterpart expression
the melting and annealing temperatures are given by $T_i=(t_{i+1}+t_{i})/2$ for any $i = 0,1,\dots, n-1$.

From the expressions $Q_{IP_S}$ and $Q_{IP_T}$  we may derive the analogue ones for the prior-posterior quadrivial
($Q_{PP_T}$ and $Q_{PP_S}$) by substituting the importance densities $g(\bs{\theta}|\,\mii)$
with the corresponding priors $\pi(\bs{\theta}|\,\mii), ~(i=0,1)$. The quadrivial expressions, univariate and multivariate,
are under ongoing research and it is not yet clear to the authors which applications could benefit from their
complected structure. The optimal tempering scheme is also an open issue. In the next section, all estimators
discussed here are applied in  simulated examples.

\section{Illustrative Examples}
 \label{sims_ch4}

\subsection{Regression modelling in the pine dataset}
For the illustration of the estimators discussed in Section \ref{sec_Estimators}
we implement the pine data set, which has been studied by \cite{FrielPet08} and \cite{Lefetal2010}
in the context of path sampling. The dataset consists of measurements taken on 42 specimens of Pinus radiata.
A linear regression model was fitted for the specimen's maximum compressive strength ($y$),
using their density ($x$) as independent variable, that is
\begin{eqnarray}\label{model}
y_i=\alpha+\beta(x_i-\bar{x})+\epsilon_i,~\epsilon_i \sim N(0,\sigma^2), ~i=1,...,42.
\end{eqnarray}
\noi
The objective in this example is to illustrate how each method and path
combination responds to prior uncertainty. To do so, we use three different prior schemes, namely:
\begin{itemize}

  \item [] $\Pi_1:~~ (\alpha, \beta)' \sim N\left\{(3000,185)', (10^6,10^4)'\right\}$,  $\sigma^2 \sim IG(3,1.8 \times 10^5)~$,
  \item [] $\Pi_2:~~ (\alpha, \beta)' \sim N\left\{(3000,0)', (10^5,10^3)'\right\}$,  $\sigma^2 \sim IG(3,1.8 \times 10^4)~$ ,
  \item [] $\Pi_3:~~(\alpha, \beta)' \sim N\left\{(3000,0)', (10^5,10^3)'\right\}$,  $\sigma^2 \sim IG(0.3,1.8 \times 10^4)$,
\end{itemize}
\noi where $IG(a,b)$ denotes the inverse gamma distribution with shape $a$ and rate $b$.
The marginal likelihoods were estimated
over $n_1=50$ and $n_2=100$ evenly spaced temperatures. At each temperature, a
Gibbs algorithm was implemented and 30,000 posterior observations were generated;
after discarding 5,000  as a burn-in period.
The posterior output was divided into 30 batches (of equal size of $R_b$=1,000 points)
and all estimators were computed within each batch.
The mean over all batches was used as the final estimate, denoted by $\log \widehat{\lambda}_{i}$ for each prior $\Pi_i$, $i=1,2,3$.
In order the estimators to be directly comparable in terms of error,
the batch means method (\citealp{Schm:82}, \citealp{bratetal:1987}) was preferred. In particular,
the standard deviation of the $\log \widehat{\lambda}$ over the 30 batches was considered
as the estimated error, denoted hereafter by $\widehat{MCE}$.
\cite{Lefetal2010} used $n=1001$ equally spaced points to compute the gold standard for
$\log \hat{\lambda}_{1}=-309.9$. Following the same approach
we derived $\log \hat{\lambda}_{2}=-323.3$ and $\log \hat{\lambda}_{3}=-328.2$. These values are considered as
benchmarks in the current study. Finally, the importance functions for each model
were constructed from the posterior means and variances at $t=1$.

\begin{table}[h!]
\caption{\small \textbf{Marginal likelihood estimates - Pine data}}
\label{Table:Likelihoods}
\small
\begin{center}
\begin{tabular} {ccccc}
\hline
\\
  $n$     & Path/Method  &  $ \log \widehat{\lambda}_{1}$   &  $\log \widehat{\lambda}_{2}$  &      $\log \widehat{\lambda}_{3}$  \\
\hline
        &PP$_{T}$ &  -312.9   ~{\scriptsize(0.21)}  & -324.7  ~{\scriptsize(0.19)} & -352.4  ~{\scriptsize(0.57)}   \\
        &PP$_{S}$ &  -310.2   ~{\scriptsize(0.06)}  & -322.6  ~{\scriptsize(0.05)} & -328.5  ~{\scriptsize(0.03)}   \\
 50     &IP$_{T}$ &  -310.0   ~{\scriptsize(0.02)}  & -323.4  ~{\scriptsize(0.03)} & -328.2  ~{\scriptsize(0.03)}   \\
        &IP$_{S}$ &  -310.0   ~{\scriptsize(0.02)}  & -323.4  ~{\scriptsize(0.03)} & -328.2  ~{\scriptsize(0.03)}   \\
\\
100     &PP$_{T}$ &  -311.3   ~{\scriptsize(0.11)}  & -323.7  ~{\scriptsize(0.14)} & -339.0  ~{\scriptsize(0.03)}   \\
        &PP$_{S}$ &  -310.1   ~{\scriptsize(0.06)}  & -323.5  ~{\scriptsize(0.03)} & -328.5  ~{\scriptsize(0.03)}   \\
        &IP$_{T}$ &  -309.9   ~{\scriptsize(0.02)}  & -323.4  ~{\scriptsize(0.02)} & -328.2  ~{\scriptsize(0.03)}   \\
        &IP$_{S}$ &  -309.9   ~{\scriptsize(0.02)}  & -323.4  ~{\scriptsize(0.02)} & -328.2  ~{\scriptsize(0.03)}   \\
\hline
 \multicolumn{5}{p{10.5cm}}{ \scriptsize{PP denotes the prior-posterior path and IP the importance posterior path.
                           The indices T and S imply the thermodynamic and stepping--stone analogues.}} \\
\end{tabular}
\end{center}
\end{table}

The estimations for the marginal  likelihoods are presented in Table \ref{Table:Likelihoods}.
The values that were obtained based on the importance-posterior path, reached the gold standards
even when $n=50$. The thermodynamic (IP$_{T}$) and the stepping--stone (IP$_{S}$) counterparts performed equally well
and were associated with similar errors.  On the contrary, the estimators that are based on the prior-posterior
path yielded different values depending on the method. In particular, the stepping--stone estimator (PP$_{S}$)
was fairly close to the gold standards with low error, for all prior schemes. The thermodynamic estimator (PP$_{T}$)
on the other hand, underestimated the marginal likelihood and exhibited higher errors than all other methods.
Logarithms of the ratios of the estimated marginal likelihoods
along with the estimated BF values directly derived by the model-switch methods
are further presented in Table \ref{Table:Ratios}.
The thermodynamic and stepping-stone analogues of MS, Q$_{PP}$ and  Q$_{IP}$,  yielded estimates with similar values and errors.

\begin{table}[h!]
\caption{\small \textbf{Estimated $\boldsymbol{\log}$ ratio of the marginal likelihoods}}
\label{Table:Ratios}
\small
\begin{center}
\begin{tabular} {lccccc}
 \hline
                        &               \multicolumn{2}{c}{ $n=50$}           & &                \multicolumn{2}{c}{ $n=100$}                  \\
                                  \cline{2-3}                                                        \cline{5-6}
                                 &&&&&\\
  Path/Method           &    $\log \left(\widehat{\lambda}_{2}/\widehat{\lambda}_{1}\right) $    & $\log \left(\widehat{\lambda}_{3}/\widehat{\lambda}_{1}\right) $   & &  $\log \left(\widehat{\lambda}_{2}/\widehat{\lambda}_{1}\right)$ &   $\log \left(\widehat{\lambda}_{3}/\widehat{\lambda}_{1}\right) $\\
\hline
      PP$_{T}$          &   -11.8 ~{\scriptsize(0.21)} &  -39.5 ~{\scriptsize(0.57)} &&  -12.4 ~{\scriptsize(0.14)} & -26.0 ~{\scriptsize(0.38)} \\
      PP$_{S}$          &   -12.5 ~{\scriptsize(0.06)} &  -18.4 ~{\scriptsize(0.73)} &&  -12.5 ~{\scriptsize(0.06)} & -18.5 ~{\scriptsize(0.34)} \\
      IP$_{T}$          &   -13.4 ~{\scriptsize(0.04)} &  -18.2 ~{\scriptsize(0.04)} &&  -13.4 ~{\scriptsize(0.03)} & -18.2 ~{\scriptsize(0.04)} \\
      IP$_{S}$          &   -13.4 ~{\scriptsize(0.04)} &  -18.2 ~{\scriptsize(0.04)} &&  -13.4 ~{\scriptsize(0.03)} & -18.2 ~{\scriptsize(0.01)} \\
      MS$_{T}$          &   -13.5 ~{\scriptsize(0.01)} &  -18.2 ~{\scriptsize(0.01)} &&  -13.5 ~{\scriptsize(0.01)} & -18.2 ~{\scriptsize(0.01)} \\
      MS$_{S}$          &   -13.5 ~{\scriptsize(0.01)} &  -18.2 ~{\scriptsize(0.01)} &&  -13.5 ~{\scriptsize(0.01)} & -18.2 ~{\scriptsize(0.01)} \\
      Q$_{PP_{T}}$      &   -13.5 ~{\scriptsize(0.01)} &  -18.2 ~{\scriptsize(0.01)} &&  -13.5 ~{\scriptsize(0.01)} & -18.2 ~{\scriptsize(0.01)} \\
      Q$_{PP_{S}}$      &   -13.5 ~{\scriptsize(0.01)} &  -18.2 ~{\scriptsize(0.02)} &&  -13.5 ~{\scriptsize(0.01)} & -18.2 ~{\scriptsize(0.01)} \\
      Q$_{IP_{T}}$      &   -13.5 ~{\scriptsize(0.01)} &  -18.2 ~{\scriptsize(0.01)} &&  -13.5 ~{\scriptsize(0.01)} & -18.2 ~{\scriptsize(0.01)} \\
      Q$_{IP_{S}}$      &   -13.5 ~{\scriptsize(0.01)} &  -18.2 ~{\scriptsize(0.01)} &&  -13.5 ~{\scriptsize(0.01)} & -18.2 ~{\scriptsize(0.01)} \\
\hline
 \multicolumn{6}{p{15cm}}{ \scriptsize{PP denotes the prior-posterior path and IP the importance posterior path.
                            MS and Q stand for the model-switch and quadrivial (bidirectional) methods.
                            The indices T and S imply the thermodynamic and stepping--stone analogues.}} \\

\end{tabular}
\end{center}
\end{table}

In this example, we have used a uniform temperature schedule, moderate number of points $n$ and non informative priors.
It was therefore reasonable to expect that the prior-based methods would be associated with higher error.
The interesting result here was that the stepping--stone estimator addressed the prior uncertainty
more successfully. In fact, the thermodynamic and stepping--stone approaches coincided only when the
gold standard was reached, which means that the discretisation error (\ref{CG_bias}) was minimized.
The next step in our analysis was to employ a  temperature schedule that places more points towards the prior
in order to reduce the uncertainty. The powered fraction (\ref{schPrior}) schedule \citep{FrielPet08}
was used with $\mathcal{C}=5$. For $n=100$, the PP$_T$ yielded the benchmark values
for the marginal likelihoods, namely  $\log \hat{\lambda}_{1}=310.0 ~(0.01)$, $\log \hat{\lambda}_{2}=323.5 ~(0.01)$
and $\log \hat{\lambda}_{2}=328.3 ~(0.02)$. The results were almost identical for  the PP$_S$.

Once the thermodynamic procedure yielded the benchmark values, we proceeded with the estimation of
the entropy measures (see Section \ref{subsec_NTI}) presented in Table \ref{Table:Entropy}. The
precision for the point $t^*$ was set  to the third decimal point and the extra MCMC runs
costed less than a minute of computational time. The Bhattacharyya and Bhattacharyya-–Hellinger values
indicate that the priors $\Pi_1$,  $\Pi_2$ and $\Pi_3$ where very distant from the corresponding posteriors.
On the contrary, the importance functions were close approximations of their matching posterior
densities. This fact completely explains the differences in the estimation,
reflecting the increased local variances encountered by the PP$_T$ as opposed to IP$_T$.

\begin{table}[h!]
\small
\caption{\small \textbf{Estimated $f-$divergencies}}
\label{Table:Entropy}
\begin{center}
\begin{tabular} {lcccccccc}
 \hline
 \\
                                      &                  \multicolumn{2}{c}{ $\Pi_1$}                       &&                       \multicolumn{2}{c}{ $ \Pi_2$}             &&       \multicolumn{2}{c}{ $ \Pi_3$}                \\
                                               \cline{2-3}                                                                                \cline{5-6}                                            \cline{8-9}
                                 &&&&&&&&\\
     \textbf{$f-$divergency}          &  PP$_T$                           &  IP$_T$                        & &        PP$_T$                   &  IP$_T$                       &&         PP$_T$                               &  IP$_T$     \\
\hline
     $ KL~(p_1 \parallel p_0)  $      &   5.6   ~{\scriptsize($<$0.01)} &   0.03 ~{\scriptsize($<$0.01)}  &&   16.3 ~{\scriptsize($<$0.01)}   & 0.10 ~{\scriptsize($<$0.01)}  &&   24.8 ~{\scriptsize($<$0.01)}      & 0.10 ~{\scriptsize($<$0.01)}    \\
     $ KL~(p_0 \parallel p_1) $       &   414.8 ~{\scriptsize(4.61)}    &   0.06 ~{\scriptsize($<$0.01)}  &&  304.1 ~{\scriptsize(5.71)}      & 0.09 ~{\scriptsize($<$0.01)}   &&   3061.0 ~{\scriptsize(53.1)}        & 0.09 ~{\scriptsize($<$0.01)}    \\
     $ J~(p_0, p_1)  $                &   420.5 ~{\scriptsize(4.62)}    &   0.09 ~{\scriptsize($<$0.01)}  &&  320.4 ~{\scriptsize(5.63)}      & 0.20 ~{\scriptsize($<$0.01)}   &&   3085.0 ~{\scriptsize(53.4)}        & 0.02 ~{\scriptsize($<$0.01)}    \\
     $ Bh~(p_0, p_1) $                &   2.53  ~{\scriptsize($<$0.01)} &   0.01 ~{\scriptsize($<$0.01)}  &&   6.68 ~{\scriptsize($<$0.01)}   & 0.03 ~{\scriptsize($<$0.01)}  &&   11.4 ~{\scriptsize($<$0.01)}      & 0.07 ~{\scriptsize($<$0.01)}    \\
     $ He~(p_0, p_1) $                &   0.96  ~{\scriptsize($<$0.01)} &   0.11 ~{\scriptsize($<$0.01)}  &&   0.99 ~{\scriptsize($<$0.01)}   & 0.17 ~{\scriptsize($<$0.01)}  &&   0.99 ~{\scriptsize($<$0.01)}      & 0.26 ~{\scriptsize($<$0.01)}    \\
     $ C_{t^*}~(p_0\parallel p_1) $   &   3.38  ~{\scriptsize($<$0.01)} &   0.01 ~{\scriptsize($<$0.01)}  &&   7.24 ~{\scriptsize($<$0.01)}   & 0.03 ~{\scriptsize($<$0.01)}  &&   15.0 ~{\scriptsize($<$0.01)}   & 0.03 ~{\scriptsize($<$0.01)}    \\
     $ R_{t^*}~(p_0\parallel p_1) $   &   2.76  ~{\scriptsize($<$0.01)} &   0.01 ~{\scriptsize($<$0.01)}  &&   4.61 ~{\scriptsize($<$0.01)}   & 0.02 ~{\scriptsize($<$0.01)}  &&   12.1 ~{\scriptsize($<$0.01)}   & 0.02 ~{\scriptsize($<$0.01)}    \\
     $ T_{t^*}~(p_0\parallel p_1) $   &   1.19  ~{\scriptsize($<$0.01)} &   0.02 ~{\scriptsize($<$0.01)}  &&   1.57 ~{\scriptsize($<$0.01)}   & 0.06 ~{\scriptsize($<$0.01)}  &&   1.24 ~{\scriptsize($<$0.01)}   & 0.06 ~{\scriptsize($<$0.01)}    \\
\hline
     $ t^*$                           &   0.183                         &   0.552                         &&   0.445                          & 0.363                         &&   0.192                          & 0.437                                 \\
\hline
 \multicolumn{9}{p{16.5cm}}{ \scriptsize{ $KL(\cdot \parallel\cdot)$: Kullback-Leibler relative entropy,
                                        $J(\cdot , \cdot)$: Jeffreys' divergence,
                                        $Bh(\cdot ,\cdot)$: Bhattacharyya distance,
                                        $He(\cdot ,\cdot)$: Bhattacharyya–-Hellinger distance. Estimated at $t^*$:
                                        $C(\cdot \parallel \cdot)$: Chernoff information,
                                        $R(\cdot \parallel\cdot)$: R\'{e}nyi relative entropy,
                                        $T(\cdot \parallel \cdot)$: Tsallis relative entropy. PP denotes
                                        the prior-posterior path and IP the importance posterior path.
                                        The indices T and S imply the thermodynamic and stepping--stone analogues. }} \\
\end{tabular}
\end{center}
\end{table}

\subsection{Marginal likelihood for latent trait models in a simulated dataset}

According to our results, the uncertainty in the pine data example was manageable
under a suitable tempering schedule. This will not always be the case, especially in high dimensional problems.
Here we consider also a factor analysis model with binary items.
The dataset consists of $N=400$ responses, $p=4$ observed items and $k=1$ latent variable
and was previously considered in \cite{Vit2013}, within the context of marginal likelihood estimation.
Under a non informative prior for the 404 model parameters (see \citealp{Vit2013} for details regarding the model specification)
the marginal likelihood was estimated close to -977.8, based on a modification of the \cite{chib:jel06} estimator
and the Laplace-Metropolis (\citealp{LewRaf97}) estimator. Using the same prior and importance functions as
in  \cite{Vit2013}, we applied the PP and the IP paths, to derive the estimated marginal likelihood.
Due to the dimensionality of the model, $n=200$  runs were used  and 30,000 posterior observations
from a Metropolis within Gibbs algorithm were derived at each temperature point
(burn in period: 10,000 iterations, thinned by 10).

The batch means for the thermodynamic and stepping-stone importance posteriors were $-978.1$ and $-977.9$ respectively,
with associated MCE errors 0.018 and 0.013. The corresponding values under the prior posterior path were
$-995.4$ and $-995.1$ with associated MCE errors 0.032 and 0.027 respectively. The low MCEs indicated that
the error was not stochastic but rather due to the temperature placement. Even though the powered fraction (\ref{schPrior}) schedule
was used to place more values close to the prior ($\mathcal{C}=5$), the uncertainty was not successfully addressed. The
estimators did not improve when the process was replicated for $n=500$. This example indicates that in high dimensional models
with non informative priors, the PP$_T$ and PP$_S$ estimators can be deteriorated by discretisation error even for large $n$.


\section{Discussion}
 \label{ch_4_disc}

In this paper we have started our quest from general thermodynamic approaches using geometric paths,
we passed from the normalized thermodynamic integration to f-divergencies,
and, finally, concluding to marginal likelihood and Bayes factors estimators.

Our study through these topics offers
 a direct connection between thermodynamic integration and divergence measures such as
Kullback-Leibler and Chernoff divergencies, Chernoff information and other divergencies
emerging as special cases or functions of them.
By this way, we were  able to offer an efficient MCMC based thermodynamic algorithm for the estimation
of the Chernoff information for a general framework which was not available in the past.

Moreover, the study of the thermodynamic identities and integrals has lead us to an understanding of the error sources of the TI estimators.
All these are accompanied with detailed graphical and geometric representation and interpretation offering
insight to the thermodynamic approach of estimating ratios of normalizing constants.

Finally, we have focused our attention on the most popular implementation of thermodynamic integration in Bayesian statistics:
the estimation of the marginal likelihood and the Bayes factors.
We have first presented an alternative thermodynamic approach based on the stepping-stone identity
introduced in biology by \cite{Xieetal11} and \cite{Fanetal11}.
By this way, we were able to present in parallel the available in the literature estimators
under the two different approaches (thermodynamic and stepping-stone) and
further introduce new appropriate estimators (based on equivalent paths)
filling in the blanks in the list  of the marginal likelihood and Bayes factors estimators.
We have also introduced quadrival Bayes factor estimators which are based on nested, more complex, paths
which seem to perform efficiently when estimating directly Bayes factors instead of marginal likelihoods.

The unified framework in thermodynamic integration presented in this article offers new highways
for research and further investigation.
Here we discuss only some of the possible future research directions.

The first one is the identification of a possible link
between the deviance information criterion, DIC, \citep{Spieg02} and thermodynamic integration.
It is well-known that in mixture models there are problems in estimating the number of efficient parameters.
A possible connection between TI and DIC may offer alternative ways of estimating it in
cases with multimodal posterior densities.
The connection between TI and KL as well as the connection between AIC, DIC and KL leave promises
that such a connection can be achieved.

A second research direction is the development of a stochastic TI approach
where the temperature will be treated as a unknown parameter.
In this case, a suitable prior should be elicitated in order to a-priori support
points where higher uncertainty of $\widehat{\mathcal{E}}_t$ is located.
Such a stochastic approach will eliminate the discretisation error which is an important
source of variability for TI estimators.

Finally, MCMC samplers used for Bayesian variable selection is another interesting area of implementation of the TI approach.
In such cases, interest may lie on the estimation of the normalizing constants over the whole model
space and the direct estimation of posterior inclusion probabilities of each covariate.
This might be extremely useful in large spaces with high number of covariates
where the full exploration of the model space is infeasible due to its size
and due to the existence of multiple neighborhoods of local maxima placed around well-fitted models.

%
%
%


 \bibliographystyle{apacite}
\bibliography{phdlit}

\end{document}